# Absence of long wavelength nematic fluctuations in LiFeAs

Adrian Merritt[1], Jose Rodriguez-Rivera[2], Yu Li[3], Weiyi Wang[3], Chenglin Zhang[3], Pengcheng Dai[3], Dmitry Reznik[1*]

[1]Department of Physics, University of Colorado-Boulder, Boulder, CO 80309, USA
[2]NIST Center for Neutron Research, National Institute of Standards and Technology, Gaithersburg, Maryland 20899, USA
[3]Department of Physics, Rice University, Houston, TX 77005, USA
*Corresponding Author: Dmitry.Reznik@colorado.edu

Abstract: We investigated long-wavelength nematic fluctuations in an Fe-based superconductor LiFeAs near $\mathbf{q}=(0.05,0,0)$ by measuring temperature-dependent renormalization of acoustic phonons through inelastic neutron scattering. We found that the phonons have conventional behavior, as would be expected in the absence of electronic nematic fluctuations. This observation implies that either electron-phonon coupling is too weak to see any effect or that nematic fluctuations are not present.

Just like the cuprates, iron pnictides and iron chalcogenites become superconducting when antiferromagnetic "parent" compounds are doped[1]. The parent compounds as well as low-doped superconducting ones undergo magnetic and structural phase transitions at temperatures $T_N$ and $T_s$ respectively. As a function of decreasing temperature, the system first becomes paramagnetic orthorhombic phase below $T_S$, and then orders antiferromagnetically below $T_N$ with $T_N \leq T_S$. A lot of recent research focused on the role of electronic nematicity, where electronic properties of the system occur preferentially along one of two otherwise degenerate perpendicular directions in the paramagnetic orthorhombic state.[2] Spin nematic order occurs when magnetic fluctuations break the C4 rotational symmetry but magnetic order does not form[3]. Magnetoelastic coupling[4,5] of the lattice to nematic fluctuations is responsible for orthorhombic distortion at zero and low doping. This spin-phonon coupling changes some phonon frequencies across the magnetic ordering transition, but the effect is much smaller than predicted.[6]

Nematic fluctuations, which are neglected in mean field approaches, have an additional effect on phonons as well as on the shear modulus $C_{66}$. Resonant ultrasound and 3-point bending experiments[7] showed that $\mathbf{q}=0$ nematic fluctuations soften $C_{66}$. The transverse acoustic phonon branch dispersing in the [0 1 0]- direction that connects to the **q**=0 shear mode has the same symmetry as nematic fluctuations of the same wavevector. Therefore, buildup of these fluctuations should also soften (reduce frequency) of these phonons.

Anomalous renormalization of these small q transverse acoustic phonons was reported in Ref.[8] based on inelastic x-ray scattering measurements. More precise neutron scattering experiments demonstrated that phonons soften from about 300K to $T_N$ and then harden sharply following a curve reminiscent of an order parameter. This behavior relates to the previously reported anomaly in the elastic constant[9,10] and indicates spin-phonon coupling to electronic nematic fluctuations. A quantitative study has shown that the gradual onset of the softening from far above $T_N$ scales with the volume of correlated magnetic domains, whereas the hardening below $T_N$ scales with the static order parameter.[11] Close correlation with magnetic properties and the similarity of the softening with the temperature-dependence of $C_{66}$ provided strong evidence that these phonons couple to nematic magnetic fluctuations.

In underdoped $BaFe_{2-x}Co_xAs_2$ where the structural transition, $T_s$, appears above $T_N$ the phonon softening correlates with the former. Thus the softening is directly related to the nematic phase. These underdoped samples showed an additional, entirely unexpected effect: Strong phonon softening below the superconducting transition, $T_c$.[12] Similar measurements on optimally-doped and overdoped samples where there is no structural or magnetic transition showed that the softening below $T_c$ reflects a buildup of nematic fluctuations at nonzero wavevectors on cooling that begins far above $T_c$. This behavior at small nonzero wavevectors is unexpected, as nematic fluctuations at **q**=0 are suppressed by superconductivity. According to the conventional theory, this competition with superconductivity should carry over to small nonzero wavevectors, but it does not.

Here we report neutron scattering results for a large high quality single crystal of LiFeAs, which comes from a different (so-called 111) family of Fe-based superconductors. This material is superconducting even when it is not doped, and it remains tetragonal at all temperatures. For this work we report **q**, Q in reciprocal lattice units (r. l. u.) with a lattice constant of 3.77 Å.

The experiment was performed on the BT-9 MACS spectrometer at the NIST Center for Neutron Research (NCNR) using one analyzer channel. Effectively, the spectrometer was used as a cold triple-axis instrument. Measurements were performed in the vicinity of the [2 0 0] Bragg peak with $E_i$ fixed at 8meV. A PG filter reduced higher order neutrons. We collected good statistics at 300K, 20K, and 1.4K; note that 20K is just above the superconducting transition temperature $T_c$=19K. The sample was a high quality single crystal of LiFeAs, whose preparation is described elsewhere[13]. For the neutron measurements we followed the experimental procedure of Ref. (9).

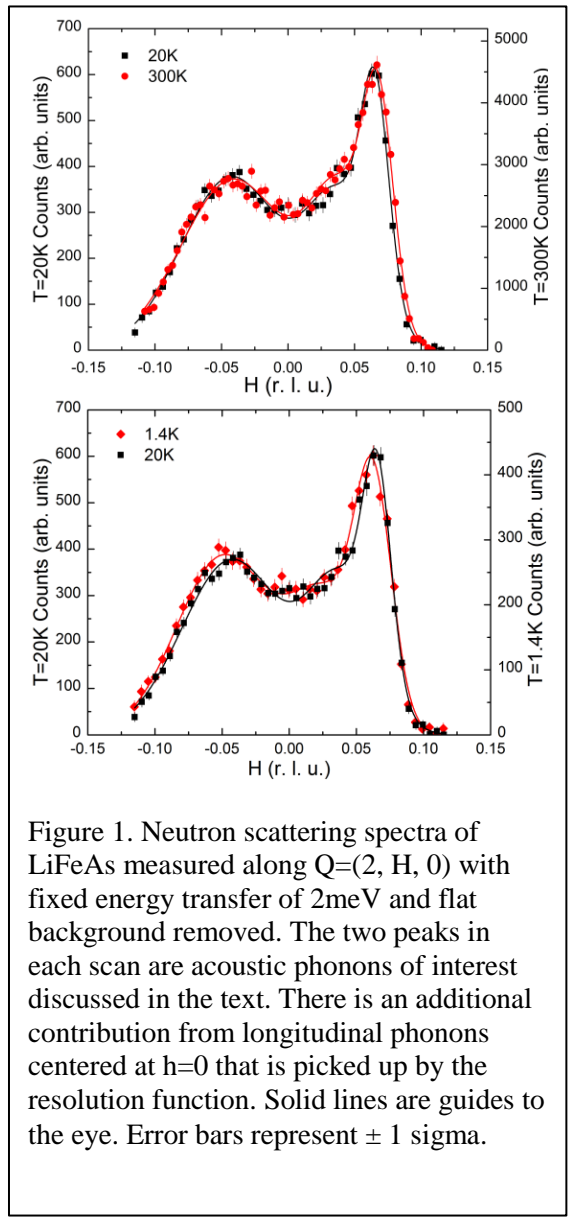

Figure 1. Neutron scattering spectra of LiFeAs measured along Q=(2, H, 0) with fixed energy transfer of 2meV and flat background removed. The two peaks in each scan are acoustic phonons of interest discussed in the text. There is an additional contribution from longitudinal phonons centered at h=0 that is picked up by the resolution function. Solid lines are guides to the eye. Error bars represent ± 1 sigma.

Figure 1 shows the results. The top panel comparing data at 300K and 20K shows almost the same peak positions; the two peaks seem to move slightly closer together on cooling by approximately 3%. This means that the phonon dispersion becomes steeper, i.e. there is hardening as opposed to softening observed in the 122 system. The bottom panel shows that the data at 20K and 3K are essentially the same. In both cases it is clear that no significant softening is occurring.

A small hardening from 300K to 20K is expected from conventional theory and is consistent with the reduction in the bond lengths as the lattice contracts on cooling. If nematic fluctuations build up on cooling, then softening of the phonon is expected. In this case the peaks in Fig. 1 would move further apart. We conclude that phonons in LiFeAs do not show any signatures of nematic fluctuations, either because the nematic fluctuations are not present or because spin-phonon coupling is weak.

Our results are consistent with the incommensurability of magnetic fluctuations ($\delta$~0.07) reported previously[10]. Furthermore, magnetic neutron scattering showed that the spectral weight of magnetic fluctuations is smaller in LiFeAs than in the 122 compounds[14]. Both of these observations can explain why nematic fluctuations in LiFeAs are not observed. It is important to perform the measurements at **q**=0 by resonant ultrasound or 3-point bending to establish the **q**=0 behavior. Based on our results, we expect that these measurements will show no signature of nematic fluctuations as well.

D.R. would like to thank R. Fernandez and J. Schmalian for helpful discussions. Y.L. would like to thank X.C. Wang for helpful discussion about crystal growth. A.M. and D.R. were supported by DOE, Office of Basic Energy Sciences, Office of Science, under Contract No. DE-SC0006939. Work at Rice is supported by DOE, BES DE-SC0012311 (P.D.) and in part by Robert A. Welch Foundation Grant No. C-1839 (P.D.). This work utilized facilities supported in part by the National Science Foundation under Agreement No. DMR-1508249

[1] P. C. Dai, RMP **87**, 855 (2015)

[2] Editors, Wang, N. L, Hosono, H, Dai, P. C., "Materials, properties, and mechanisms of iron-based superconductors", ISBN: 978-9-81430-322-4.

[3] X.Y. Lu, J. Park, R. Zhang, H. Luo, A. Nevidomskyy, Q. Si, P. C. Dai Science 345, 657 (2014)

[4] L. Boeri, O. V. Dolgov, and A. A. Golubov, Phys. Rev. Lett. **101**, 026403 (2008).

[5] Z. P. Yin, S. Lebegue, M. J. Han, B. Neal, S. Y. Savrasov and W. E. Pickett, Phys. Rev. Lett. **101**, 047001 (2008).

[6] D. Reznik, K. Lokshin, D. C. Mitchell, D. Parshall, W. Dmowski, D. Lamago, R. Heid, K.-P. Bohnen, A. S. Sefat, M. A. McGuire, B. C. Sales, D. G. Mandrus, A. Subedi, D. J. Singh, A. Alatas, M. H. Upton, A. H. Said, A. Cunsolo, Yu. Shvyd'ko, and T. Egami, Phys. Rev. B **80**, 214534 (2009).

[7] Fernandes RM, Chubukov AV, & Schmalian J (2014) What drives nematic order in iron-based superconductors? Nature Physics **10**(2):97-104.

[8] J. L. Niedziela, D. Parshall, K. A. Lokshin, A. S. Sefat, A. Alatas, and T. Egami, Phys. Rev. B **84**, 224305 (2011)

[9] R. M. Fernandes, A. E. Böhmer, C. Meingast, and J. Schmalian, " Scaling between Magnetic and Lattice Fluctuations in Iron Pnictide Superconductors," Phys. Rev. Lett. **111**, 137001 (2013)

[10] A. E. Böhmer, P. Burger, F. Hardy, T. Wolf, P. Schweiss, R. Fromknecht, M. Reinecker, W. Schranz, and C. Meingast, Phys. Rev. Lett. **112**, 047001 (2014).

[11] D. Parshall, L. Pintschovius, D. Lamago, J.-P. Castellan, J. L. Niedziela, Th. Wolf, D. Reznik Phys. Rev. B **91**, 134426 (2015).

[10] N. Qureshi, P. Steffens, Y. Drees, A. C. Komarek, D. Lamago, Y. Sidis, H.-J. Grafe, S. Wurmehl, B. Büchner, and M. Braden Phys. Rev. Lett. **108**, 117001 (2012)

[12] F. Weber, D. Parshall, L. Pintschovius, J.-P. Castellan, M. Merz, Th. Wolf, and D. Reznik, arXiv:1610.00099 [cond-mat.supr-con] (2016)

[13] L. Y. Xing, H. Miao, X.C. Wang, J Ma, Q. Q. Liu, Z. Deng, H. Ding, and C. Q. Jin, J. Phys.: Condens. Matter **26,** 435703 (2014)

[14] Y. Li, Z. Yin, D. W. Tam, D. L. Abernathy, A. Podlesnyak, C. Zhang, M. Wang, L. Xing, C. Jin, K. Haule, G. Kotliar, T. A. Maier, P. C. Dai, Phys. Rev. Lett. **116**, 247001 (2016)